\documentclass[conference]{IEEEtran}
\IEEEoverridecommandlockouts
\usepackage{cite}
\usepackage{amsmath,amssymb,amsfonts}
\usepackage{algorithmic}
\usepackage{graphicx}
\usepackage{textcomp}
\usepackage[many]{tcolorbox}
\usepackage{xcolor}
\def\BibTeX{{\rm B\kern-.05em{\sc i\kern-.025em b}\kern-.08em
    T\kern-.1667em\lower.7ex\hbox{E}\kern-.125emX}}
\usepackage{enumitem}
\setlist{nolistsep}
\usepackage{booktabs}
\usepackage{multirow}
\usepackage{tabularx}
\usepackage{tablefootnote}
\usepackage{siunitx}
\usepackage{etoolbox}
\definecolor{main}{HTML}{373737}
\definecolor{sub}{HTML}{c5c6d0}

\newtcolorbox{titledBox} {
    space to upper,
    skin=bicolor,
    colbacklower=main,
    collower=sub,
    halign=left,
    valign=center,
    halign lower=flush right
}

\AfterEndEnvironment{enumerate}{\vskip-\lastskip}

\begin{document}


\title {Evolution of Phishing Detection with AI: A Comparative Review of Next-Generation Techniques}
\title{Phishing Detection in the Gen-AI Era:\\ Quantized LLMs vs Classical Models}


\author{\IEEEauthorblockN{ Jikesh Thapa, Gurrehmat Chahal, Șerban Voinea Gabreanu, Yazan Otoum  \\
\IEEEauthorblockA{{\textit{School of Computer Science and Technology}}}
\textit{Algoma University, Canada}\\
Emails: \{jithapa, gurchahal, svoineagabreanu, otoum\}@algomau.ca}\vspace{-2em}
}
\IEEEoverridecommandlockouts

\maketitle

\IEEEpubidadjcol

\begin{abstract}

Phishing attacks are becoming increasingly sophisticated, underscoring the need for detection systems that strike a balance between high accuracy and computational efficiency. This paper presents a comparative evaluation of traditional Machine Learning (ML), Deep Learning (DL), and quantized small-parameter Large Language Models (LLMs) for phishing detection. Through experiments on a curated dataset, we show that while LLMs currently underperform compared to ML and DL methods in terms of raw accuracy, they exhibit strong potential for identifying subtle, context-based phishing cues. We also investigate the impact of zero-shot and few-shot prompting strategies, revealing that LLM-rephrased emails can significantly degrade the performance of both ML and LLM-based detectors. Our benchmarking highlights that models like DeepSeek R1 Distill Qwen 14B (Q8\_0) achieve competitive accuracy, above 80\%, using only 17~GB of VRAM, supporting their viability for cost-efficient deployment. We further assess the models' adversarial robustness and cost-performance tradeoffs, and demonstrate how lightweight LLMs can provide concise, interpretable explanations to support real-time decision-making. These findings position optimized LLMs as promising components in phishing defence systems and offer a path forward for integrating explainable, efficient AI into modern cybersecurity frameworks.

\end{abstract}


\section{Introduction}

Phishing is a prevalent cyberthreat that manipulates users into divulging sensitive information, such as credentials and financial data. Between November 2023 and January 2024, the Cybercrime Information Center collected approximately one million phishing reports \cite{putra2024analysis}. With the increasing sophistication of phishing tactics, traditional detection approaches struggle to keep pace with evolving attack techniques. Phishing has become more challenging in the 2020s due to the increased use of social media and multi-vector attacks that combine various methods, including email, text messages, and social media platforms. They have evolved from simple, deceptive emails to highly sophisticated social engineering schemes. Attackers leverage Artificial Intelligence (AI) generated content, deepfake techniques, and multi-channel deception strategies to exploit human vulnerabilities. Integration of AI, particularly Machine Learning (ML) and Natural Language Processing (NLP), has shown promising results in alleviating some of this burden. Several recent studies published between 2021 and 2024 have demonstrated the effectiveness of deep learning models, including Convolutional Neural Networks (CNNs) and Recurrent Neural Networks (RNNs) \cite{otoum2024advancing}, in detecting phishing attempts with higher accuracy compared to classical approaches, with XGBoost achieving the highest accuracy at 99.89\%, followed by PILFER with 99.5\% \cite{alghenaim2025state}. However, recent advancements in Large Language Models (LLMs) and Generative AI (GenAI) \cite{otoum2025llm, otoum2025llms} further exacerbate the threat, enabling automated and highly targeted phishing campaigns. The authors of \cite{afane2024nextgen} found that while traditional phishing detectors, such as Gmail Spam Detector, SpamAssassin, Proofpoint, and State-of-the-Art LLMs perform well on original phishing emails, their accuracy and recall decline notably when dealing with LLM-rephrased versions of the duplicate emails. The study displayed a significantly better classification performance from LLMs with rephrased emails, and even with the original emails, the performance was still marginally more accurate. Unfortunately, the performance gains of LLMs come at a significant cost in terms of energy and computational resources. The total energy consumption for training a transformer model with 6B parameters to completion is estimated to be around $103.5 MWh$ \cite{dodge2022carbon}, and given that the computational cost of training an LLM depends on the number of parameters, modern alternatives like GPT-1, GPT-2, and GPT-3 featuring 117 million, 1.5 billion, and 175 billion parameters, respectively, the question arises whether the marginal performance gain is worth it. And if and how the LLMs or Traditional approaches can be evolved to provide a more optimal solution. This paper presents a structured review of such next-generation phishing detection techniques and their cost-to-performance ratio, focusing on the following contributions:

\begin{itemize}
    \item A comparative analysis of traditional, Machine Learning, Deep Learning, and LLM-based phishing detection approaches.
    \item Exploration of adversarial phishing tactics, LLM rephrasing, and AI/LLM-driven countermeasures.
    \item Comparative analysis of model performances and marginal performance gain against supplemental time and resource consumption.
    \item Identification of research gaps and future directions in phishing detection.
\end{itemize}

The remainder of this paper is organized as follows: Section~\ref{Literature Review} reviews related works across ML, DL, and LLM-based phishing detection, identifying key research gaps. Section~\ref{Proposed Model} presents the proposed models and methodologies, including experimental setup and dataset details. Section~\ref{CONCLUSION} summarizes the findings, highlights cost-performance trade-offs, and outlines future research directions for robust, efficient phishing detection.

\section{Literature Review}
\label{Literature Review}

\subsection{Related Works}

Phishing detection has evolved from traditional ML/NLP methods to deep learning models, and recently to LLM-driven approaches. Studies such as \cite{bethany2024llm} and \cite{kulkarni2024robustness} highlight how LLMs both amplify phishing sophistication and enhance detection under adversarial conditions. Recent work also integrates LLMs into Security Awareness Training (SAT)~\cite{is2024llm} and cybersecurity policy formulation~\cite{quinn2024applying}, reducing phishing susceptibility. The authors of~\cite{roy2024chatbots} proposed early malicious prompt detection transferable across major LLMs, achieving high accuracy, up to 96\%. Despite initial successes, conventional ML/NLP approaches~\cite{kyaw2024systematic, thakur2023sys} remain inadequate against evolving phishing attacks, motivating the development of hybrid solutions that combine LLM contextual reasoning with the proven robustness of traditional methods~\cite{otoum2025open}.

\subsection{Comparative Work}

\subsubsection{Adversarial Prompting: Using LLMs to Subvert Detection Models}
\begin{figure}[h]
  \centering
  \includegraphics[width=0.5\textwidth]{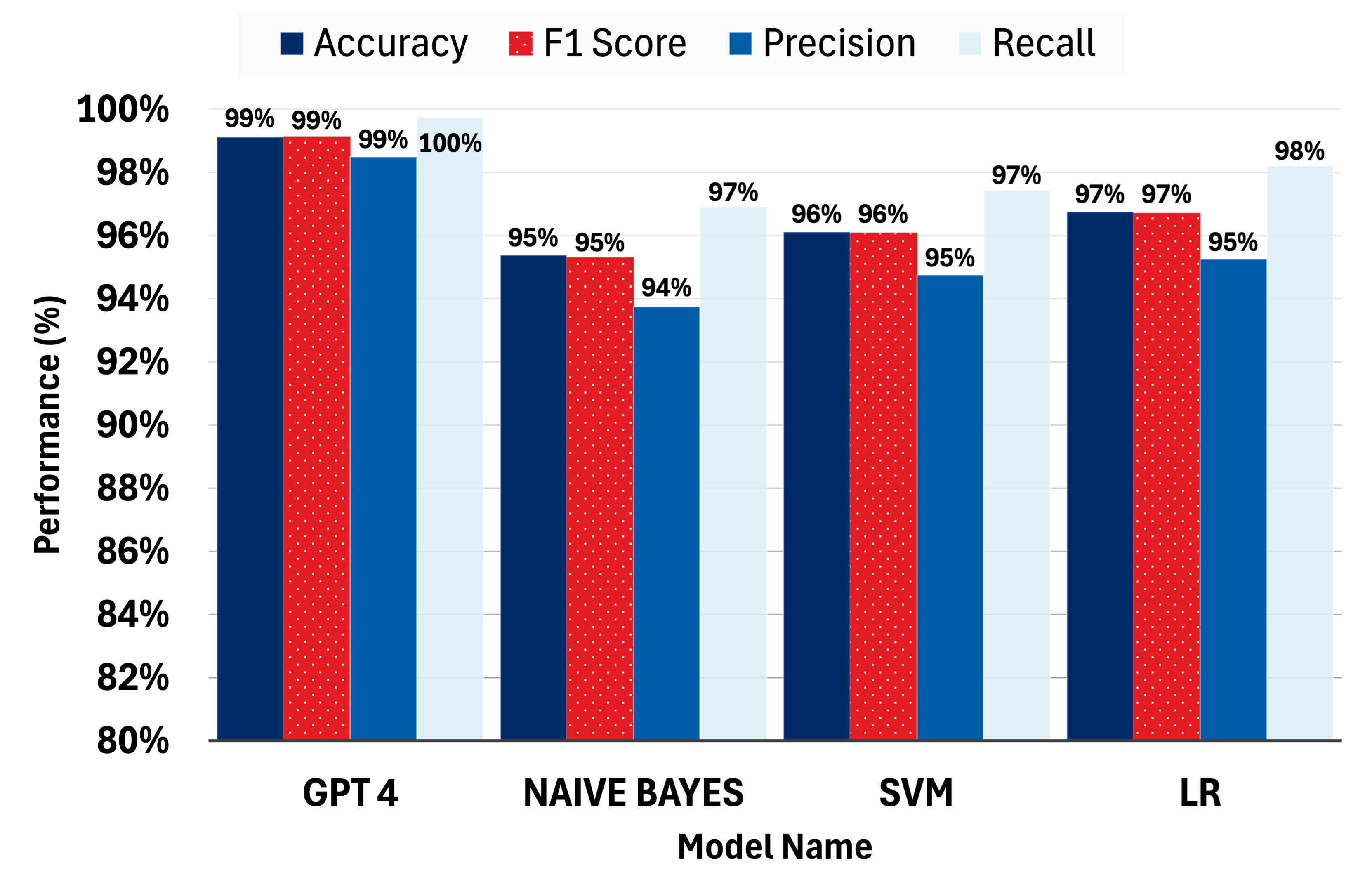}
  \caption{Comparative results of phishing‑detection models on original emails.}
  \label{fig:orig}
\end{figure}

Using LLMs to reword emails effectively reduces the efficiency of phishing detectors, with traditional Machine Learning models losing a significant amount of accuracy. LLMs such as GPT‑4, however, were shown to be able to retain more detection accuracy despite the reworded emails \cite{afane2024nextgen}. As shown in Fig.~\ref{fig:orig}, all models achieve very high accuracy when classifying the unmodified (“original”) emails. GPT‑4 tops the list at around 98.5\% accuracy, while traditional models like Naive Bayes and Logistic Regression are just behind at 97.2\% and 96.8\%, respectively.

\begin{figure}[h]
  \centering
  \includegraphics[width=0.5\textwidth]{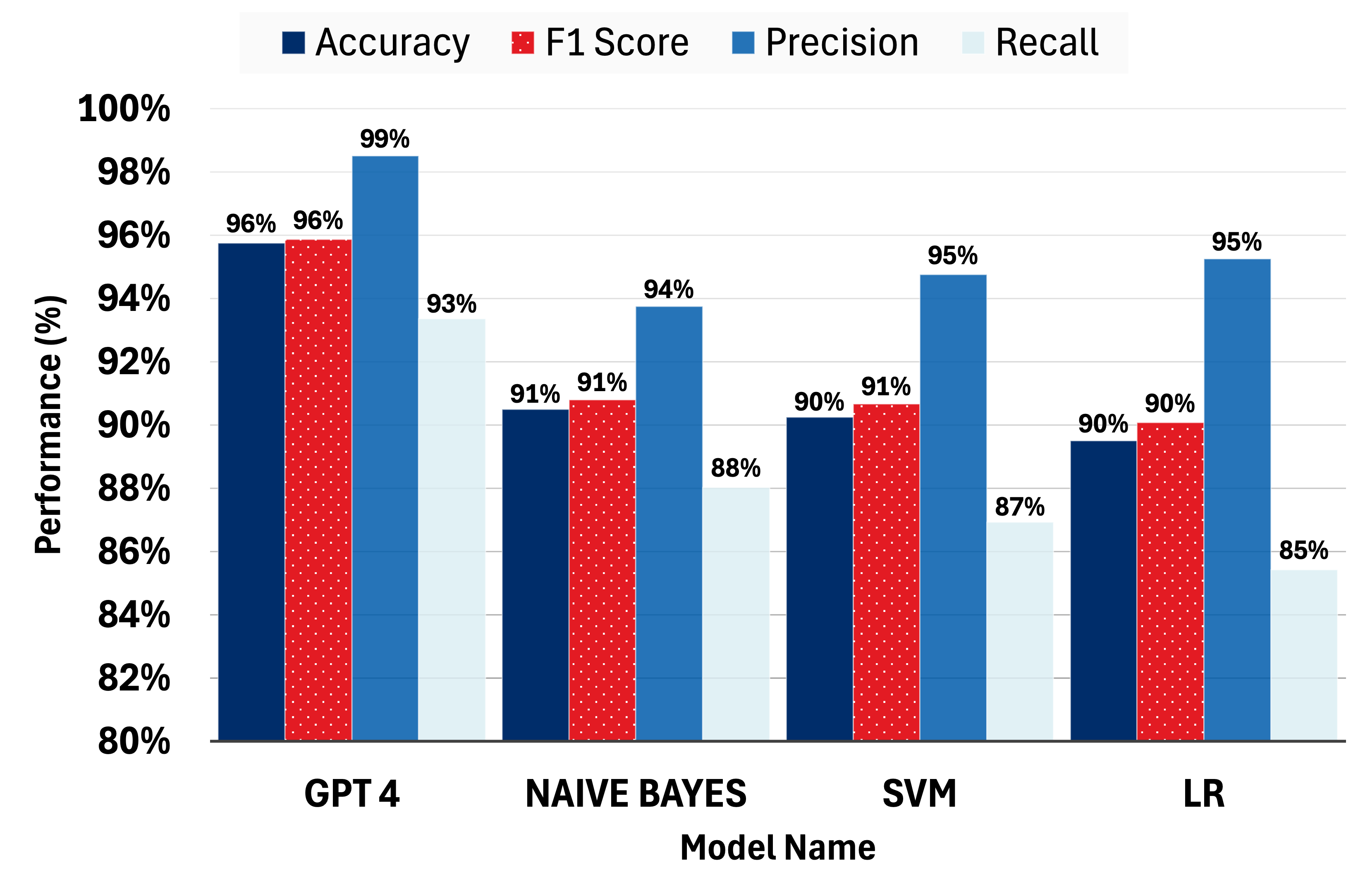}
  \caption{Comparative results of phishing‑detection models on zero‑shot rephrased emails.}
  \label{fig:zero}
\end{figure}


As illustrated in Fig.~\ref{fig:zero}, zero-shot rephrasing of phishing emails leads to a noticeable decline in detection performance across all evaluated models. Specifically, Naive Bayes exhibits a drop of approximately 5.3 percentage points, Logistic Regression declines by 6.1 percentage points, and GPT-4 experiences a reduction of about 3.2 percentage points in accuracy. This performance degradation underscores the vulnerability of both traditional and LLM-based detectors to simple LLM-generated paraphrasing techniques.

\begin{figure}[h]
  \centering
  \includegraphics[width=0.5\textwidth]{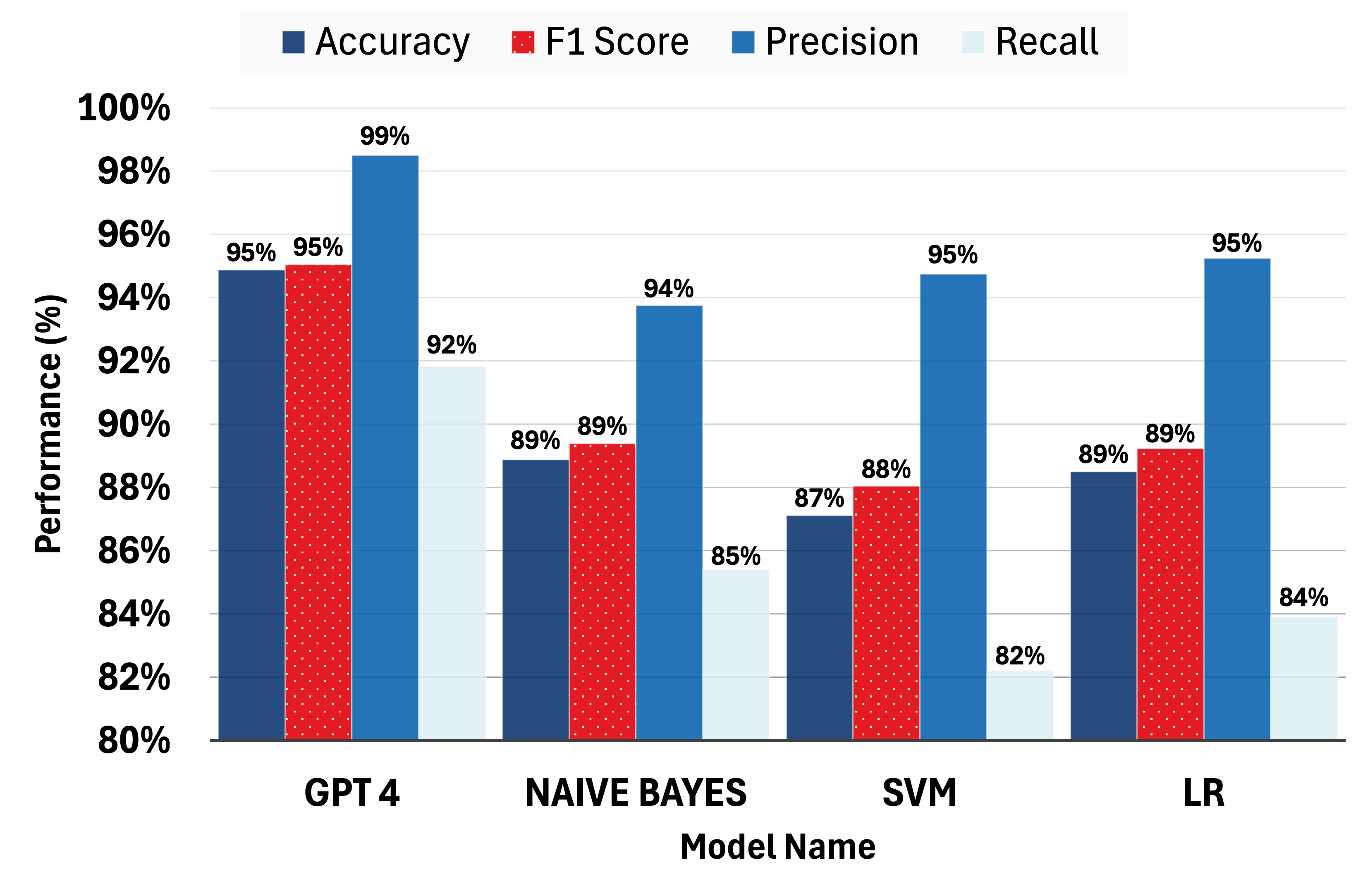}
  \caption{Comparative results of phishing‑detection models on few‑shot rephrased emails.}
  \label{fig:few}
\end{figure}


Fig.~\ref{fig:few} illustrates the impact of few-shot rephrasing on phishing detection performance, revealing a more pronounced decline than zero-shot rephrasing. Naive Bayes accuracy drops by 6.5 percentage points, Logistic Regression by 8.25 points, and SVM by 9.0 points. In contrast, GPT-4 demonstrates greater resilience, with a relatively modest decrease of 4.24 percentage points. These results highlight the comparative robustness of LLM-based detectors against paraphrased inputs. However, the significant degradation observed across traditional models highlights their limitations when confronted with advanced LLM-based evasion strategies, such as reflection and beam search, which effectively preserve malicious intent while obfuscating linguistic patterns.

\begin{itemize}
    \item Debnath et al. \cite{debnath2022email} achieved 99.14\% accuracy using BERT on the Enron Email Dataset
    \item Ali et al. \cite{ali2025text} achieved an overall classification accuracy of 99\% on the Lingspam dataset, retaining 96\% accuracy against adversarial samples
    \item Shahrivari et al. (2020) \cite{shahrivari2020phishing} demonstrated the effectiveness of various machine learning techniques for phishing detection
\end{itemize}

The research collectively indicates that while traditional ML approaches can be highly effective against conventional phishing attempts, they require significant enhancement to counter emerging LLM-generated threats. The defensive algorithm proposed by Fairbanks demonstrated a 97\% improvement in mitigating these sophisticated attacks. This comparative analysis highlights the importance of developing specialized defenses against LLM-generated phishing, as traditional methods may be insufficient against these increasingly sophisticated threats.

\subsubsection{Traditional ML Models Performance}
The “David versus Goliath” paper by Greco et al. \cite{greco2024supporting} demonstrates that smaller, traditional machine learning models can be highly effective for detecting LLM-generated phishing emails. Their experimental results show that neural networks achieved the highest accuracy at 99.78\% but with the highest resource cost, closely followed by SVM (99. 2\%) and logistic regression (99. 03\%), while other models like KNNs (97.67\%), Random Forest (98.16\%) and Naïve Bayes (94.1\%) showed relatively lower accuracy but much stronger performance. These results are further reinforced by our performance benchmarks. The authors concluded that Logistic Regression provides the best balance of accuracy, performance, and explainability making it the best fit for practical implementations.

\subsubsection{Deep Learning Models Performance and Optimization}

Deep learning (DL) models offer superior capabilities in extracting complex patterns from data but are inherently more resource-intensive than traditional machine learning methods. Their success in cybersecurity applications, including phishing detection, has led to the development of numerous DL-based solutions~\cite{altwaijry2024dlmodels}. Models such as BERT, DistilBERT, and ANN variants have demonstrated high accuracy levels across datasets like Enron and Lingspam, often exceeding 98\%. However, these gains come at the cost of significantly larger model sizes, with baseline BERT models containing over 100 million parameters. Given the emphasis on cost-efficiency in this study, we deliberately avoided using large pre-trained models. Our approach instead utilized lightweight architectures totaling 4.7 million parameters, offering a substantial reduction in computational demand while maintaining strong predictive performance. This design choice was inspired by the work~\cite{altwaijry2024dlmodels}, which demonstrated that augmenting a 1D-CNNPD model with LSTM, GRU, and their bi-directional variants could achieve high accuracy without excessive resource requirements. Our methodology builds on these findings to strike a balance between performance, scalability, and deployment feasibility.

\subsubsection{Small Vs Large LLMs}

The results below, comparing the small LLMs to Claude, show that there is significant room for improvement for the small LLMs. While not a 1-1 comparison due to different testing parameters, it still shows that with an accuracy of up to 80\%, small LLMs are capable of understanding if emails have phishing content. An LLM trained specifically on phishing datasets would understand phishing emails much better, as its parameters would be trained on phishing data rather than irrelevant information, such as the ability to code.

\begin{figure}[h]
  \centering
  \includegraphics[width=0.5\textwidth]{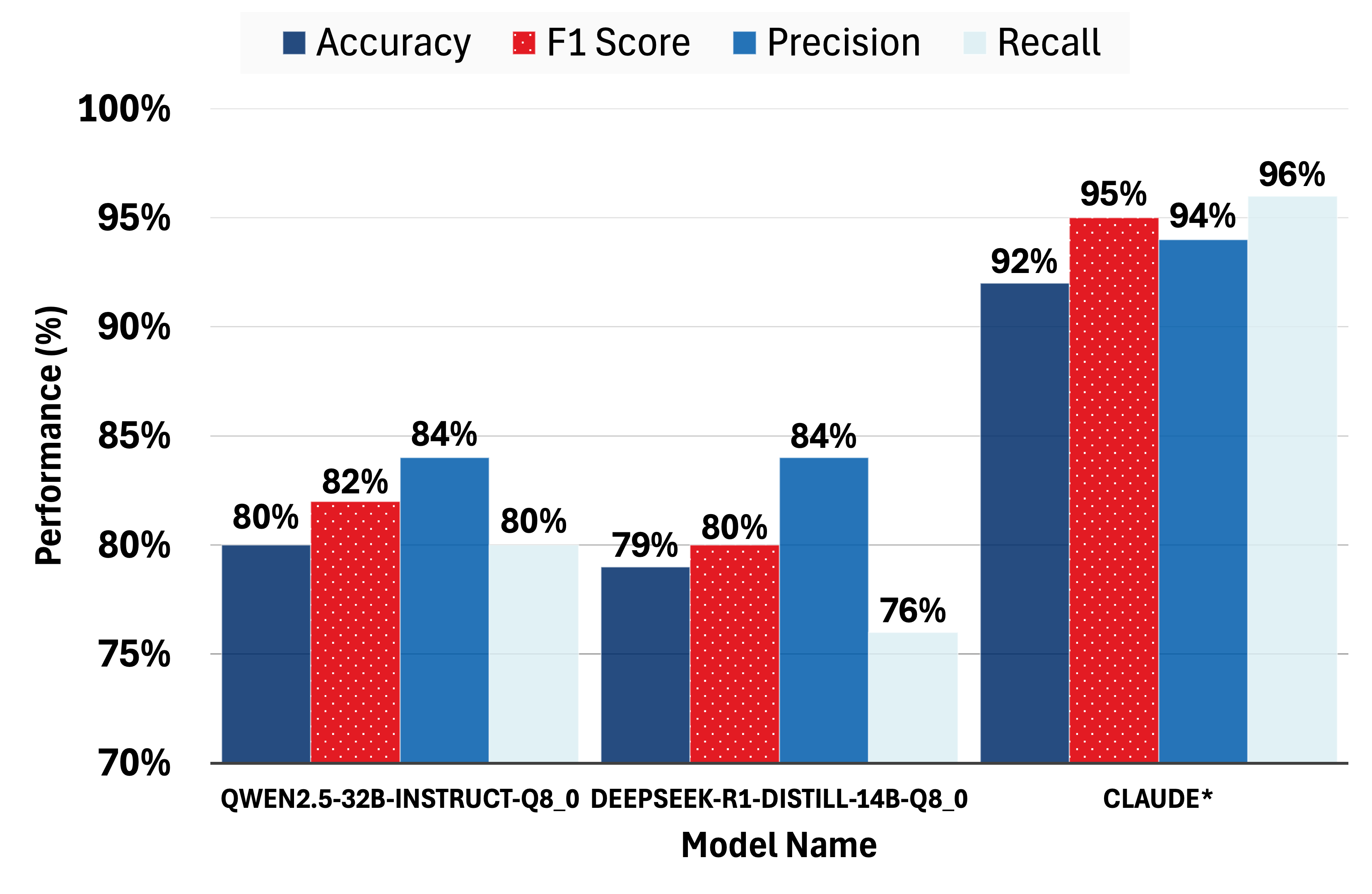}
  \caption{Performance comparison between Full \& Small LLMs.  
    The bar marked with $^*$ indicates the result for Claude taken from \cite{roy2024chatbots}.}
  \label{fig:llmperf}
\end{figure}

As shown in Fig.~\ref{fig:llmperf}, quantized small LLMs can get close to a large model’s phishing‐detection performance.  For example, Qwen 2.5 32B Q8 achieves 81\% accuracy, which is only 12\% below the significantly larger Claude model. Although these results are obtained from slightly different test sets, they still demonstrate that a modern, carefully tuned LLM would be capable of delivering a highly accurate phishing detection model while requiring fewer resources.

\subsection{Current Research Gaps}

Despite notable progress in ML, DL, and LLM-driven phishing detection, several critical gaps persist that limit the robustness, adaptability, and real-time deployment of these approaches.

\subsubsection{Lack of Specialized Phishing Detection Models}
Current LLMs are general-purpose and not optimized for phishing-specific patterns. Training LLMs exclusively on phishing-related data could enhance sensitivity to subtle deception tactics such as spear phishing and lateral attacks~\cite{bethany2024llm}.

\subsubsection{Deployment Inefficiencies and Quantization Needs}
Large LLM-based phishing models are resource-intensive, which hinders their deployment in real-time and at the edge. Quantizing models into smaller, energy-efficient variants is crucial to balance performance and operational cost~\cite{bethany2024llm}.

\subsubsection{Outdated and Limited Datasets}
Existing phishing datasets are often static, brand-specific, and fail to capture evolving attack vectors. Updated datasets, incorporating LLM-generated phishing samples and adversarial variants, are essential for developing resilient models~\cite{kulkarni2024robustness}.

\subsubsection{Incomplete Hybridization of Detection Techniques}
Standalone LLMs could be more effective when integrated with traditional rule-based or ML classifiers. Hybrid architectures can leverage the adaptability of LLMs and the precision of classical methods to improve resilience against evolving threats~\cite{kulkarni2024robustness}.

\subsubsection{Limited Research on Adversarial Robustness}
The adversarial vulnerabilities of phishing detection LLMs remain underexplored. Investigations into adversarial training, red-teaming, and watermarking techniques are crucial for enhancing their reliability in attack scenarios~\cite{bethany2024llm}.

\subsubsection{Insufficient Early Detection of Malicious Prompts}
Recent studies show promise in detecting malicious prompts that manipulate LLMs into generating phishing content~\cite{roy2024chatbots}. Early intervention mechanisms can prevent automated phishing attacks and reduce the need for complex downstream detection, substantially improving system efficiency.

\section{Proposed Model}
\label{Proposed Model}
\subsection{Methodologies}
\subsubsection{Small LLMs}
The evaluation system for testing small LLMs is implemented using a Python script that loads a subset of phishing emails from the dataset into a CSV file. It automatically discovers the GGUF-formatted LLM checkpoint with the MODEL\_FILES mapping. Using llama-cpp-python, each model is initialized with the appropriate number of GPU layers and an 8,000-context window. The script has two versions, for the thinking and non-thinking models. The thinking models can complete their thinking portion, which may take some time due to the number of tokens it requires. In contrast, the normal models are only given 100 tokens to respond with, which is enough for a True or False response and a brief explanation. For the tests conducted in this paper, the LLM's explanation is not required, and thus, the LLMs are cut off early to save time during testing, which is part of the reason the thinking models have longer run times in our tests. The script then calls the LLM and provides a prompt with instructions and the email's body. 

\begin{tcolorbox}[
    title={Test Prompt Text}
]
You are an AI email security assistant specializing in phishing detection. Your task is to analyze the email below and decide whether it is a phishing or scam attempt.

"Please follow these strict instructions:"

"1. Your response must be a single line starting with
either 'TRUE:' or 'FALSE:'"

"2. 'TRUE:' means you believe the email is a phishing/scam
attempt; 'FALSE:' means you do not."

"3. Immediately after the colon, provide a concise
explanation for your decision (do not include any extra text or commentary)."

"For example, if the email asks you to click a suspicious link for account verification, you might respond:"

"TRUE: The email contains urgent language and a suspicious
link, typical of phishing attempts."

"Now, analyze the following email:"

"Email Content: ..."
\end{tcolorbox}%

Metrics such as total generation time, time to first token, and tokens per second are recorded at the end of every response. Using a regular expression (regex) parser, the first valid classification of True or False is identified and recorded. If none are found, then the response is considered "Failed." Every result is saved periodically to a CSV file to track the raw data for each response. When testing is completed, the scikit-learn library calculates two metric sets: one that includes all failures and one that excludes them, including accuracy, precision, recall, F1 Score, and AUC-ROC. Once the program is completed, the remaining raw data is appended to the raw data CSV file, and the final results are saved as a normal text file. All the LLM tests were run using Python 3.13.2 with the following libraries llama\_cpp\_python 0.3.8, pandas 2.2.3, and scikit-learn 1.6.1. 

\subsubsection{Machine Learning}
The model evaluation pipeline was implemented using Python with libraries such as Pandas, NumPy, scikit-learn, and Matplotlib. The dataset was imported from the phishing email dataset CSV file, labels were binarized, and the data was split into training and testing sets. Text inputs were vectorized using TF-IDF with a 1000-feature limit and English stop-word removal. Logistic Regression, Naive Bayes, Random Forest, and SVM were trained and tested on the transformed data. The models were evaluated using accuracy, precision, recall, F1-score, AUC, and prediction time per 1000 emails. ROC curves were plotted for each model. The more robust the model in accuracy and speed, the more suitable it was for real-time phishing detection scenarios.

\subsubsection{Deep Learning}
Almost all the libraries used in the Machine Learning experimentation were also used in the Deep Learning approach. The same dataset and a relatively similar preprocessing pipeline were used to ensure the ultimate outcomes were more directly comparable. The maximum token limit for vectorization was set to 20,000, and padding was introduced to create a uniform input size for the neural networks. A factory class was designed to generate the model layer definitions, and functions were developed to iteratively run the models, test them, collect model metrics, and create visualizations, such as confusion matrices and ROC curves. Finally, each model was run five times to calculate the average expected performance, thereby avoiding biased results from random spikes, as the model's intended use is for practical applications. The Keras library was used to create the RNN models, and the training was conducted on two separate platforms. The training time was relatively short, with a total training time of less than 10 minutes for the four models. Even the most resource-intensive model, the Bi-Directional LSTM model, took only approximately 200 seconds to complete training. A Second training set was conducted on Google Colab to allow comparisons with the ML models' training and inference time. Multiple rounds of optimizations were performed on the code to ensure minimal resource waste.

\subsection{Dataset \& Resources}
The dataset used for testing is a classification dataset containing 83,446 emails, sourced from the 2007 TREC Public Spam Corpus and the Enron-Spam Dataset. It includes labeled data, where 1 indicates spam and 0 indicates legitimate emails for any given body text. The combined dataset was curated and made publicly available by Purusinghvi on Kaggle \cite{purusinghvi2023email}.
All the models were trained on a MacOS Sequoia 15.3.2 system using an M4 Max with 64GB of RAM, 48GB usable as VRAM.

\subsection{Results \& Analysis}

\begin{table*}
    \centering
    \caption{Comparative Analysis of AI Models for Phishing Detection (Invalid Responses Excluded)}
    \label{tab:excluded}
    \begin{tabular}{lcccccccc}
        \hline
        \textbf{Model} & \textbf{Accuracy} & \textbf{F1 Score} & \textbf{Precision} & \textbf{Recall} & \textbf{AUC-ROC} & \textbf{Avg. TPS} & \textbf{Runtime (hr)} & \textbf{~VRAM Usage} \\
        \hline
        Qwen 2.5 7B Q4 & 0.49 & 0.34 & 0.58 & 0.24 & 0.52 & 27.94 & 0.54 & 5 GB \\
        Qwen 2.5 14B Q8 & 0.55 & 0.39 & 0.74 & 0.27 & 0.58 & 11.97 & 0.84 & 15 GB \\
        Qwen 2.5 32B Q8 & 0.81 & 0.82 & 0.84 & 0.80 & 0.81 & 6.54 & 2.67 & 34 GB \\
        DeepSeek R1 Distill Qwen 7B Q4 & 0.72 & 0.76 & 0.72 & 0.82 & 0.71 & 35.00 & 4.08 & 5 GB \\
        DeepSeek R1 Distill Qwen 14B Q8 & 0.79 & 0.80 & 0.84 & 0.76 & 0.79 & 15.16 & 5.95 & 15 GB \\
        \hline
    \end{tabular}
\end{table*}

\begin{table*}
    \centering
    \caption{Comparative Analysis of AI Models for Phishing Detection (Invalid Responses Included)}
    \label{tab:included}
    \begin{tabular}{lcccccccc}
        \hline
        \textbf{Model} & \textbf{Accuracy} & \textbf{F1 Score} & \textbf{Precision} & \textbf{Recall} & \textbf{AUC-ROC} & \textbf{Avg. TPS} & \textbf{Runtime (hr)} & \textbf{~VRAM Usage} \\
        \hline
        Qwen 2.5 7B Q4 & 0.45 & 0.31 & 0.51 & 0.22 & 0.48 & 27.00 & 0.54 & 5 GB \\
        Qwen 2.5 14B Q8 & 0.52 & 0.37 & 0.68 & 0.25 & 0.55 & 11.71 & 0.84 & 15 GB \\
        Qwen 2.5 32B Q8 & 0.71 & 0.73 & 0.76 & 0.69 & 0.71 & 6.34 & 2.67 & 34 GB \\
        DeepSeek R1 Distill Qwen 7B Q4 & 0.63 & 0.68 & 0.65 & 0.71 & 0.62 & 33.83 & 4.08 & 5 GB \\
        DeepSeek R1 Distill Qwen 14B Q8 & 0.75 & 0.76 & 0.81 & 0.72 & 0.76 & 14.94 & 5.95 & 15 GB \\
        \hline
    \end{tabular}
\end{table*}

\subsubsection{Small LLM Performance}

Across the tests, the quantized small LLMs showed a wide level of performance variation. The DeepSeek R1 Distill Qwen models outperformed their same‑size non‑thinking Qwen 2.5 counterparts, achieving up to 79\% accuracy in the 14B variant, versus the 55\% accuracy for the Qwen 2.5 14B. This result suggests that small thinking LLM models may offer substantially better performance for phishing detection compared to their normal counterparts. However, they also take longer to output an answer, as shown in the runtime variation between the two types of models. This underscores that higher parameter counts do not necessarily mean a substantially better result. When the invalid responses are also included as wrong, the drop in performance is substantial in all models, but the 14B DeepSeek R1 Distill Qwen model had a lower drop in accuracy at 4\%, compared to Qwen 32B’s drop of 10\%. These results show that even without fine-tuning existing models, small LLMs such as DeepSeek R1 Distill Qwen 7B Q4\_K\_M are capable of detecting phishing emails at a higher rate than guessing, suggesting there is considerable room for improvement in a specially trained phishing detection model. When invalid responses are included, the other core metrics (F1 score, AUC‑ROC, and Recall) exhibit commensurate declines. The smallest models, Qwen 2.5 7B, and 14B variant, performed very poorly with accuracies as low as 49\% and 55\%, akin to flipping a coin. This poor performance is likely due to not enough phishing related data in their parameters. In comparison, the distilled DeepSeek R1 7B Q4\_K\_M model had a 72\% Accuracy, a 76\% F1 score, 72\% Precision, and 82\% Recall while only using around 5 GB of VRAM. When invalid responses were included, the model still had a higher level of accuracy at 63\%, which is a 9\% drop. The LLMs were also able to explain why an email may or may not be phishing, in addition to the TRUE/FALSE classification. For example: "FALSE: The email appears to be a legitimate weather advisory from the National Weather Service, providing information about coastal hazards and surf conditions without requesting personal information or including suspicious links." DeepSeek R1 Distill Qwen 14B, in this instance, not only correctly labeled the sample as non-phishing but also provided a good summary of why the email is likely legitimate. Providing such explanations to users by highlighting key indicators or a lack thereof can reduce the likelihood of a person being fooled by a phishing email. The resource trades offs are high, for instance Qwen 32B used around 127\% more VRAM than DeepSeek R1 Qwen 14B, but offered only a small 2\% increase in accuracy (81\% vs 79\%). While the deepseek generated tokens faster, it still ran significantly slower since it needed more tokens overall to think with 5.95 hours vs 2.67 hours, a 122.8\% increase in time. The “Thinking” models took longer to complete in our testing setup because we let them generate their full chain of thought before outputting a True/False decision. In contrast, the standard models were cut off a few tokens after their True/False response. This was done to limit runtime due to hardware and time constraints. Allowing the reasoning models to complete their internal thinking steps naturally added a significant amount of time to the testing, which explains part of the duration difference. DeepSeek R1 Distill Qwen is more capable in VRAM-constrained scenarios, as the 14B model only requires around 15 GB of VRAM, while the 32B version of Qwen uses more than double that, at around 34 GB. This shows several trade‑offs between each model that need to be considered when implementing LLMs in a phishing detection system. Overall, both the 32B Qwen and 14B DeepSeek R1 Distill Qwen models demonstrate significant potential for phishing detection, remarkably when fine-tuned to enhance their performance. The Tables~\ref{tab:excluded} and~\ref{tab:included} illustrate the same set of results from the LLMs, however Table~\ref{tab:included} includes invalid responses. When invalid outputs are excluded (Table~\ref{tab:excluded}), each model’s Accuracy and F1 score are higher—for example, Qwen 2.5 32B Q8 achieves 81\% accuracy (F1 0.82). Including those invalid responses as errors (Table~\ref{tab:included}) uniformly lowers performance. Qwen 2.5 32B Q8 drops to 71\% accuracy (F1 0.73), with similar 5 to 10 point declines seen in the other models. This demonstrates further prompt tuning, and LLM training would likely help improve the instruction-following capabilities of these models. One limitation is that the LLMs were not finetuned on the dataset, meaning the parameters aren’t optimized to find nuanced linguistic patterns. Fine tuning the models on a phishing specific dataset would likely provide substantial performance improvements and allow for a better comparison against the ML and DL models.

\subsubsection{Deep Learning Performance}

\begin{figure}[h]
\centerline{\includegraphics[width=0.5\textwidth]{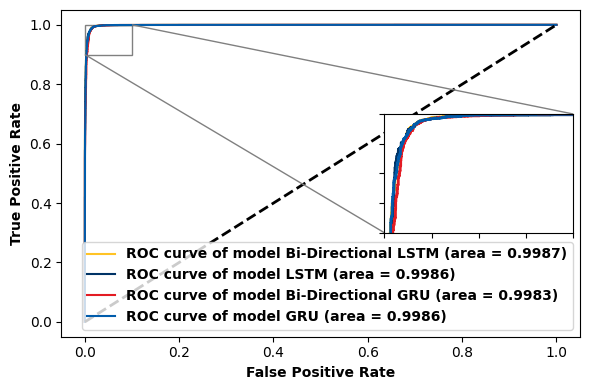}}
\caption{ROC curves for Deep Learning models}
\label{figure DL ROC Curve}
\end{figure}

Each of the proposed models is developed by augmenting the 1D-CNNPD layer with LSTM, GRU, Bi-LSTM and Bi-GRU as suggested by the work \cite{altwaijry2024dlmodels}. Additionally, the models were tested with a leaky ReLu activation in the fully connected layer, as the initial outcomes in the training epochs displayed stagnation and subsequent degradation. Note that all the results shown and plotted were generated from an average of 5 subsequent model executions and evaluations. As apparent from the ROC curve in Fig.~\ref{figure DL ROC Curve}, even the regular Deep Learning models provide a significant performance boost over regular Machine Learning models, with almost all the models displaying an AUC over 0.998. There are some marginal gains among the models as well, with Bi-Directional LSTM performing the best in terms of accuracy (98.74\%) and GRU performing the worst (98.62\%), although only slightly. Although this is acceptable, the loss metrics during the training epochs showed room for improvement.

\begin{figure}[h]
\centerline{\includegraphics[width=0.5\textwidth]{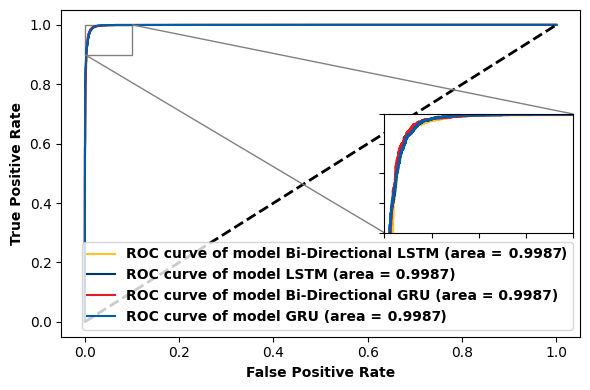}}
\caption{ROC curves for Deep Learning models with Leaky ReLU Activation}
\label{figure DL Leaky ROC Curve}
\end{figure}

During training, it was observed that performance gains began to plateau after the seventh epoch. One of the possible diagnoses was dying neurons, which is a common problem when using ReLu activation. The dying ReLU refers to a situation in which ReLU neurons become inactive due to their lower limit of 0. There are many empirical and heuristic explanations of why ReLU neurons die. However, little is known about its theoretical analysis \cite{lulu2020dyrelu}. To remedy this issue, we used the suggested Leaky ReLu activation function. As shown in Fig.~\ref{figure DL Leaky ROC Curve}, this improvement enhanced the overall model performance, particularly for GRU-based models. Table~\ref{tab:model_comparison} supports the outcome displayed in Fig.~\ref{figure DL Leaky ROC Curve}, Leaky ReLU activation provides marginal performance gains for most models. It produces the best performing model so far, the Bi-Directional GRU, with an accuracy of 98.77\% and an AUC of 0.9987 (non-rounded). As a GRU model, it also has shorter training and inference times. This model was the most optimal in terms of cost-to-performance among all the models tested.

\begin{table}[htbp]
    \centering
    \caption{Model Performance Comparison: ReLU vs Leaky ReLU Activation}
    \label{tab:model_comparison}
    \begin{tabular}{lcccc}
        \toprule
        \multirow{2}{*}{\textbf{Model}} & \multicolumn{2}{c}{\textbf{ReLU Activation}} & \multicolumn{2}{c}{\textbf{Leaky ReLU Activation}} \\
        \cmidrule(lr){2-3} \cmidrule(lr){4-5}
        & \textbf{Accuracy} & \textbf{ROC-AUC} & \textbf{Accuracy} & \textbf{ROC-AUC} \\
        \midrule
        LSTM & 0.98664 & 0.99859 & 0.98706 & 0.99872 \\
        Bi-LSTM & 0.98736 & 0.99875 & 0.98730 & 0.99866 \\
        GRU & 0.98628 & 0.99824 & 0.98748 & 0.99871 \\
        Bi-GRU & 0.98748 & 0.99857 & 0.98766 & 0.99874 \\
        \bottomrule
    \end{tabular}
\end{table}

\subsubsection{ML Performance}
\begin{figure}[h]
\centerline{\includegraphics[width=0.5\textwidth]{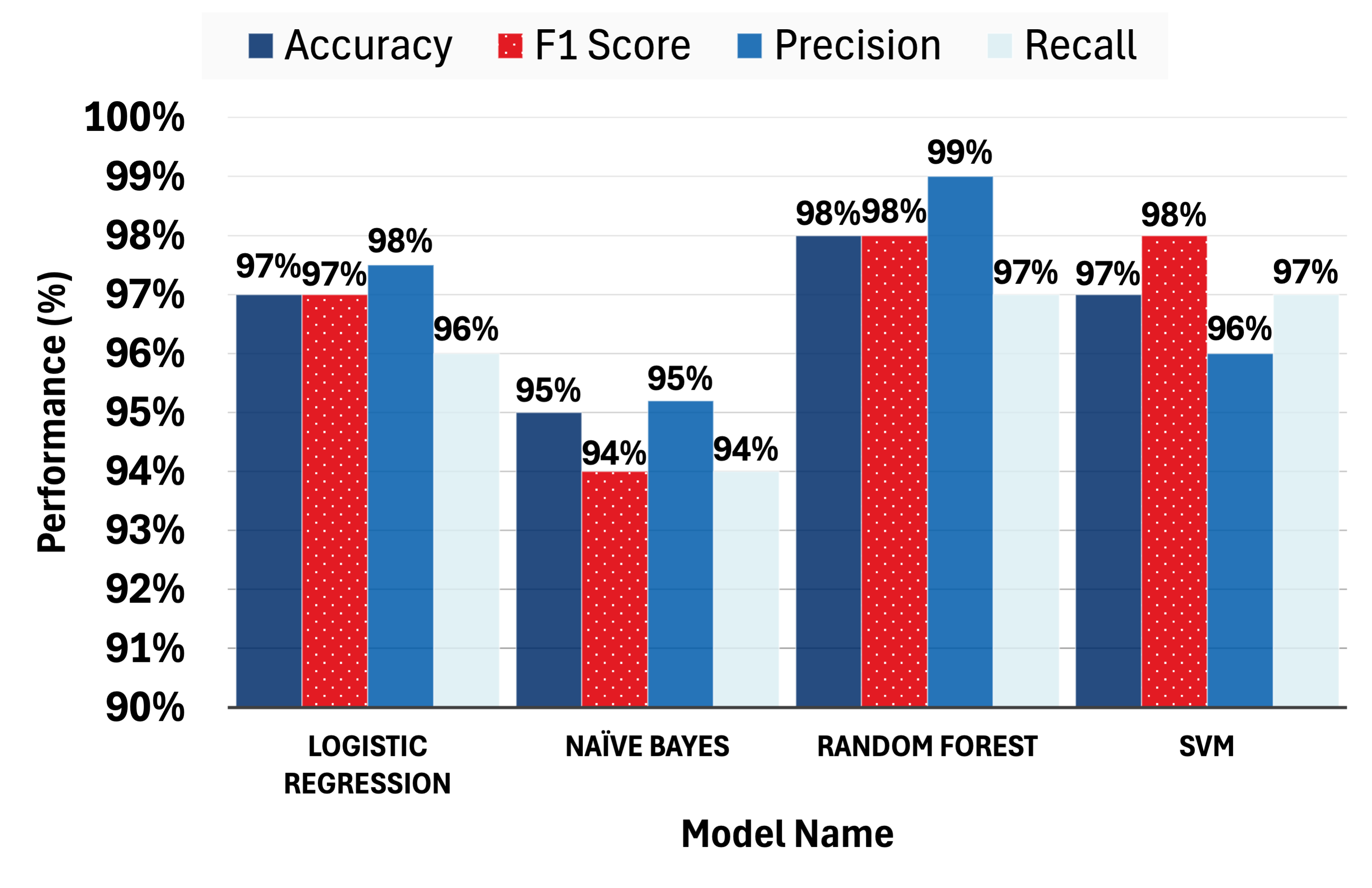}}
\caption{Comparative results of Machine Learning Algorithms}
\label{figure: ml results}
\end{figure}

\begin{figure}[h]
\centerline{\includegraphics[width=0.5\textwidth]{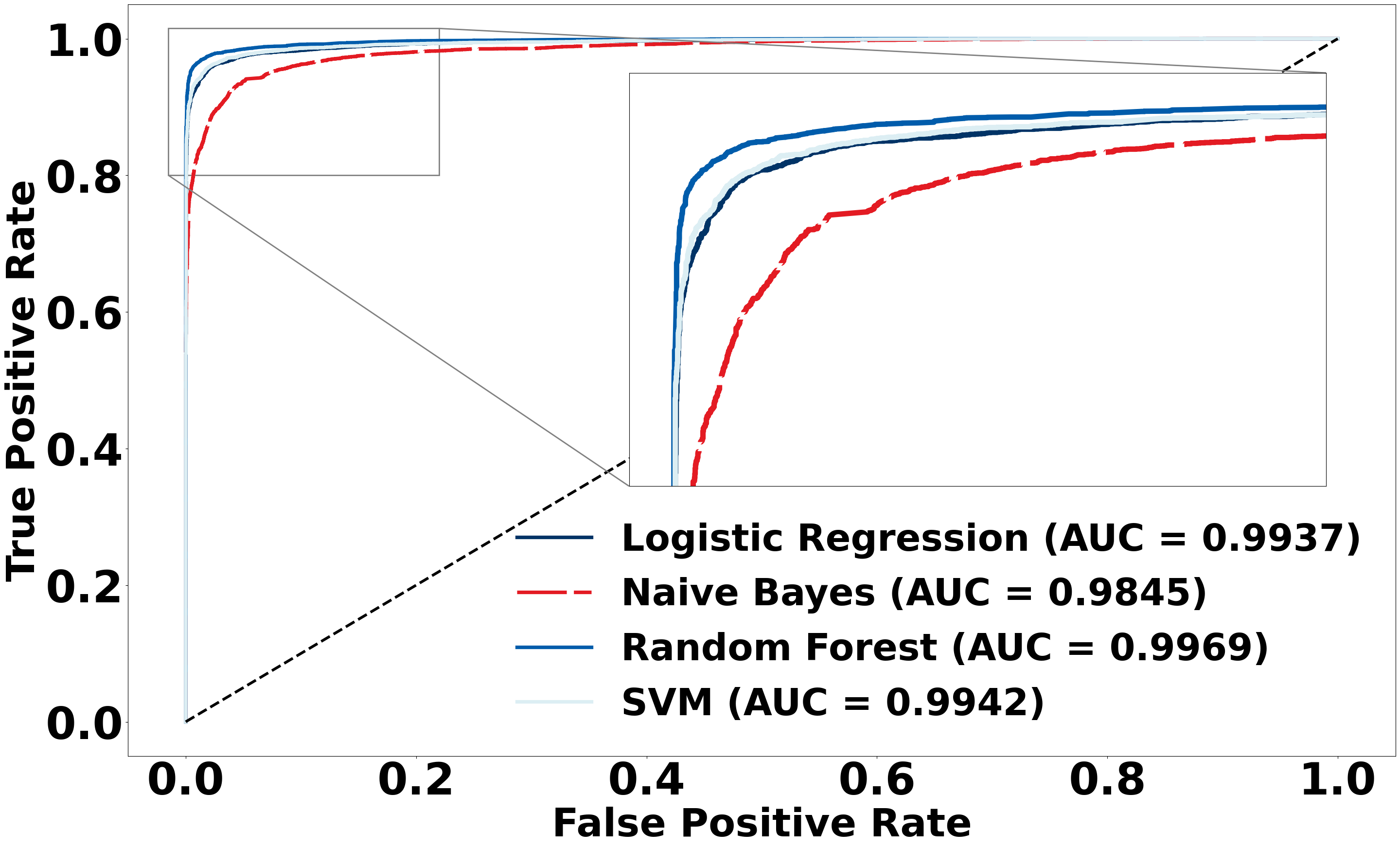}}
\caption{ROC curves for all the tested ML models}
\label{figure ML ROC Curves}
\end{figure}

Among the evaluated models for phishing detection, Random Forest stands out as having the highest overall performance in our tests. Fig.~\ref{figure: ml results} and Fig.~\ref{figure ML ROC Curves} show that it achieves top scores across all metrics, accuracy (0.9801), precision (0.9862), recall (0.9718), and F1-score (0.9789), indicating strong reliability and minimal error rates. However, its prediction time of 0.0276 seconds per 1000 emails, though reasonable, is higher compared to faster models. Logistic Regression offers a more balanced solution, combining solid performance accuracy of 0.9704, F1-score of 0.9686, and an unmatched prediction speed of 0.0001 seconds. This makes it especially suitable for real-time or large-scale deployments where speed and accuracy are essential. It remains a practical and efficient choice for phishing detection systems. Naive Bayes, with the lowest accuracy (0.9466) and F1-score (0.9435), excels in speed, processing 1000 emails in only 0.0002 seconds. It is ideal for constrained environments requiring high throughput, but can be less reliable in terms of Precision. SVM performs well in terms of accuracy (0.9711) and F1-score (0.9694); however, its major limitation is slow processing time, which takes 1.3934 seconds for 1000 emails, making it less practical for real-time applications.

In summary, Random Forest is the most accurate, Logistic Regression is the most balanced and scalable, Naive Bayes is the fastest, and SVM offers high precision but at a cost of efficiency.

\subsubsection{Resource Consumption}

One of the primary objectives of this research was to identify the overall resource consumption of each approach and the resultant performance gain. This idea was prompted by the massive reported costs of training and deploying LLMs, as well as the impact of such expenses on their applicability in practical scenarios.

\begin{table}[htbp]
    \centering
    \caption{Model Efficiency Comparison}
    \label{tab:model-comparison}
    \begin{tabular}{lll}
        \hline
        \textbf{Model Name$^a$} & \textbf{Inference Time (s)$^b$} & \textbf{Best Accuracy} \\
        \hline
        Qwen 2.5 7B Q4 & 1944 & 49\% \\
        Qwen 2.5 14B Q8 & 3024 & 55\% \\
        Qwen 2.5 32B Q8 & 9612 & 81\% \\
        DeepSeek R1 7B Q4 & 14688 & 72\% \\
        DeepSeek R1 14B Q8 & 21420 & 79\% \\
        \hline
        Logistic Regression & 0.0001 & 97\% \\
        Naive Bayes & 0.0003 & 95\% \\
        Random Forest & 0.0033 & 98\% \\
        Linear SVM & 0.0006 & 97\% \\
        \hline
        GRU & 0.3850 & 98.75\% \\
        Bi-GRU & 1.4250 & 98.77\% \\
        LSTM & 0.4400 & 98.71\% \\
        Bi-LSTM & 1.4458 & 98.74\% \\
        \hline
    \end{tabular}
\end{table}


Table~\ref{tab:model-comparison} shows the inference time for each of the LLM, ML, and DL models tested in the paper. The ML models are approximately 1308 times faster than the DL models in inference, with nearly the same performance, but with an average of 2\% less accuracy. The ML models are also approximately 1.38 million times faster than the LLMs, illustrating that using LLMs to check every email is significantly less efficient. Instead, LLMs are best suited for complex cases that require contextual reasoning and human-readable explanations. However, in order to accommodate this sophisticated functionality, LLMs consume a massive amount of precious resources such as water and energy. Moreover, the resulting carbon footprint poses a significant concern. In a comprehensive research conducted by Jegham et al. \cite{jegham2025hungryaibenchmarkingenergy}, where 30 major LLMs were analyzed, it was found that the most resource-intensive LLMs, such as ChatGPT-O3 and DeepSeek R1, can consume up to $30Wh$ per long prompt. More significantly, the research estimates that with the measured $0.43Wh$ per query and the reported 700 million queries per day, annually, ChatGPT-4o uses energy comparable to approximately 35,000 US households and evaporates freshwater matching the annual drinking needs of 1.2 million people, concluding a chicago-sized forest would be needed to offset the carbon footprint. Wong V. \cite{Wong_2024} calculated that serving a single prompt in ChatGPT produces more than 4 grams of $CO_2eq$, which is over 20 times the operational carbon footprint of a web search query.  The environmental impact becomes even more concerning when considering that findings published by Chien et al. \cite{chien2023reduce} show the energy consumption of LLM serving has now surpassed that of training. This shift toward serving-dominated energy consumption creates a complex sustainability problem that requires comprehensive solutions such as hardware lifecycle management and energy efficiency optimization as proposed by Ding et al. \cite{ding2024sustainable}. The results from our tests align well with these observations: during inference, the 14 billion–parameter DeepSeek R1 Distill Qwen Q8\_0 model required approximately 6 hours of compute and 15 GB of VRAM to identify phishing emails. In comparison, the best‑performing deep learning model Leaky ReLU Bi‑GRU achieved better performance, with an accuracy of 98.77\% versus 79\% for DeepSeek R1 Distill Qwen 14B Q8\_0 in just 15.33 seconds, substantially less than the 5.95 hours DeepSeek R1 Distill Qwen 14B Q8\_0 took. While the testing platforms were not identical, the substantial difference in performance far exceeds what could be explained by hardware alone. Classical machine learning approaches performed similarly to DL, with models such as random forest requiring only a few seconds to run and providing around 98\% accuracy. These findings suggest a hybrid approach for real‑world deployments where ML and DL models handle the bulk of cases. At the same time, a fine‑tuned small LLM intervenes in the most challenging scenarios, optimizing accuracy, cost, and environmental impact simultaneously.


\section{CONCLUSION}
\label{CONCLUSION}

This study evaluated phishing detection approaches across traditional ML, DL, and small LLMs. Our findings reveal that models like DeepSeek R1 Distill Qwen 14B achieve over 79\% accuracy using only 15~GB of VRAM, while Bi-GRU models exceed 98\% accuracy with minimal inference time, highlighting their cost-effectiveness. Despite their limitations in raw accuracy, LLMs provide contextual explanations and resilience against adversarial rephrasing, key advantages for real-time phishing detection. A hybrid framework that combines the strengths of ML/DL with context-aware LLMs can deliver improved accuracy, interpretability, and efficiency. Future work should focus on specialized LLM training using phishing-centric datasets, robust adversarial defence strategies, and real-time threat integration. This hybrid approach offers a practical path toward scalable, interpretable, and environmentally conscious phishing detection systems in the Gen-AI era. To improve future phishing detection, work should focus on developing specially trained LLMs that primarily use phishing-related content to enhance their detection of nuanced and context-driven attacks. Creating large, fully trained models for phishing detection and then quantizing them can also be done to strengthen computational requirements. The results from this paper are limited by the reliance on English-only sources in the dataset. This fails to account for linguistic and cultural diversity of global phishing attacks, incorporating modern multilingual datasets could produce a universally effective model. Finally, combining trained LLMs and traditional detection methods, such as ML, in a hybrid approach would be the most optimal overall scenario. Ultimately, focus should be on enhancing adversarial resilience and facilitating early detection of malicious prompts, thus significantly mitigating misuse while reducing computational costs.

\end{document}